\documentclass[twocolumn]{aastex6}

\epsscale{1.17} 

\usepackage{epsfig}
\usepackage{amsmath}
\usepackage{amssymb} 
\usepackage{epstopdf}

\begin{document}

\title{Polarization Signature of Companion-Fed Supernovae Arising from BH-NS/BH Progenitor Systems}

\author{Xudong Wen$^{1}$, He Gao$^{1,*}$, Shunke Ai$^{2,*}$,Liang-Duan Liu$^{3,4}$, Jin-Ping Zhu$^{5,6}$ And Wei-Hua Lei$^{7}$}
\affiliation{
$^1$Department of Astronomy, Beijing Normal University, Beijing 100875, China; gaohe@bnu.edu.cn\\
$^2$Department of Astronomy, School of Physics and Technology, Wuhan University, Wuhan 430072, China; shunke.ai@whu.edu.cn\\
$^3$Institute of Astrophysics, Central China Normal University, Wuhan 430079, China.\\
$^4$Key Laboratory of Quark and Lepton Physics (Central China Normal University), Ministry of Education, Wuhan 430079, China.\\
$^5$School of PhysicsA and Astronomy, Monash University, Clayton, VIC 3800, Australia.\\
$^6$OzGrav: The Australian Research Council Centre of Excellence for Gravitational Wave Discovery, Clayton, VIC 3800, Australia.\\
$^7$Department of Astronomy , School of Physics, Huazhong University of Science and Technology, Wuhan, Hubei 430074, China.}

\begin{abstract}
The formation of black hole-neutron star (BH-NS) or BH-BH systems may be accompanied with special supernova (SN) signals, due to the accretion feedback from the companion BH. The additional heating, which is mainly attributed to the Blandford-Payne mechanism, would disrupt the isotropic nature of the luminosity distribution on the surface of the SN ejecta, leading to the appearance of polarization. Here we develop a three dimensional (3D) Monte Carlo polarization simulation code (MCPSC) to conduct simulations for these special SNe. We find that the maximum polarization level of approximately $\sim 2\%$ occurs at the peak time of SN emission in the ``close-binary" scenario, while in the ``faraway-binary" case, maximum polarization (i.e. $\sim 0.7\%$) is observed at a considerably later time than the peak of the SN. The magnitude of polarization is dependent on the degree of unevenness in the luminosity distribution and the angle between the line of sight and the equatorial direction. When considering the geometric distortion of supernova ejecta at the same time, the magnitude of polarization may either increase (for a oblate ellipsoidal shape) or decrease (for a prolate ellipsoidal shape). The polarization signatures represent a promising auxiliary instrument to facilitate the identification of the companion-fed SNe. Moreover, by comparing the event rate of these special SNe with the event rate density of LIGO-Virgo detected BH–NS/BH systems could further help to distinguish the BH–NS/BH formation channel.

\end{abstract}

\keywords{Supernovae (1668); Polarimetry(1278);
Gravitational waves (678); Black holes (162);
}

\section{Introduction}

Since the groundbreaking detection of the first gravitational wave (GW) event, GW150914 \citep{Abbott2016}, emanating from the merger of a binary-black-hole (BH-BH) system, the LIGO and Virgo scientific collaborations, which are now complemented by KAGRA \citep{Akutsu2021}, have confirmed the occurrence of a total of 90 compact binary-system mergers during the initial three observing runs \citep{Abbott2021a,Abbott2021b}, among which are two binary-neutron-star (NS-NS) mergers (namely, GW170817 \citep{Abbott2017a,Abbott2017b} and GW190425 \citep{GW190425}), various possible black hole-neutron star (BH-NS) mergers \citep{Abbott2021a}, as well as a large number of BH-BH mergers. With the onset of the fourth observing run (O4), the LIGO-Virgo-KAGRA collaboration (LVKC) has entered a phase of regular gravitational wave observations.

At present, the formation channel of the BH-NS/BH system is still under debated. There are mainly two scenarios for the formation of BH-NS/BH systems, one of which is the isolated binary evolution in the galaxy fields \citep{tutukov73,lipunov97,belczynski16}, and the other is the dynamical interaction in dense environments \citep{sigurdsson93,portegies00,rodriguez15}. In the binary evolution scenario, the faster evolving star is first to produce a BH through core collapse, and forms a BH - massive star binary. The second core collapse event after a certain time delay will then result in the formation of the BH-NS/BH system.

\cite{Gao2020} proposed that the formation of a BH-NS/BH system through isolated binary evolution might be accompanied by a special supernova due to the accretion process of the first-formed companion BH (henceforth companion-fed SN). 
In this scenario, the companion BH injects additional energy into the supernova ejecta through the Blandford-Payne (BP) mechanism \citep{BP}, resulting in a sharp peak in the lightcurve with luminosity even up to the level of superluminous supernovae (SLSNe), or a plateau feature compared to the regular luminosity of core collapse SNe. The non-spherical injection of energy from the BP mechanism imparts a heterogeneous luminosity distribution on the photosphere surface of the supernova ejecta, with the highest concentration of energy observed in the equatorial direction. This phenomenon has the potential to induce polarization signals,
and the origin of these polarization signals differs from the classical cause of polarization, which arise from a non-spherical photosphere \citep{Shapiro82,Hoflich91,Hoflich96,Dessart11,Bulla15}, the blocking effect of absorbing material above the photosphere \citep{Kasen03,Hole10,Tanaka2017}, and the existence of an off-center radiation source \citep{Hoflich95}. 

In this work, we develop a 3-dimensional (3D) Monte Carlo polarization simulation code (MCPSC) to calculate the continuum polarization properties of the companion-fed SNe. The primary objective is to advance the identification of such kind of distinctive SNe through polarization observations in the future, leading to a better comprehension of the generation channel of the BH-NS/BH systems.

\section{Methods}
\label{sec:methods}

\subsection{Physical model} 

Consider a binary system consisting of a massive star and a companion BH (with mass $M_{\mathrm{BH}}$) with an orbital separation $d$. As the massive star explodes as a SN, a total mass $M_{\rm ej}$ with an explosive energy $E_{\mathrm{sn}}$ is ejected.
We assume that the SN ejecta undergoes an homologous expansion i.e., $r = v t$ and the density profile of SN ejecta follows a broken power law \citep{matzner99}, which can be expressed as

\begin{equation}
  \rho_{\mathrm{ej}} (v, t) = \left\{ \begin{array}{ll}
    \zeta_{\rho} \frac{M_{\mathrm{ej}}}{v_{\mathrm{tr}}^3 t^3} \left(
    \frac{r}{v_{\mathrm{tr}} t} \right)^{- \delta}, & v_{\mathrm{ej}, \min}
    \leqslant v < v_{\mathrm{tr}}\\
    \zeta_{\rho} \frac{M_{\mathrm{ej}}}{v_{\mathrm{tr}}^3 t^3} \left(
    \frac{r}{v_{\mathrm{tr}} t} \right)^{- n}, & v_{\mathrm{tr}} \leqslant v
    \leqslant v_{\mathrm{ej}, \max}
  \end{array} \right.
\end{equation}
where the transition velocity $v_{\mathrm{tr}}$ could be obtained from the density continuity condition
\begin{equation}
\begin{aligned}
  v_{\mathrm{tr}} &= \zeta_v \left( \frac{E_{\mathrm{sn}}}{M_{\mathrm{ej}}}
  \right)^{1 / 2} \\ \nonumber
  &\simeq 1.2 \times 10^4 \mathrm{km} \text{ s}^{- 1}  \left(
  \frac{E_{\mathrm{sn}}}{10^{51} \mathrm{erg}} \right)^{1 / 2} \left(
  \frac{M_{\mathrm{ej}}}{M_{\odot}} \right)^{- 1 / 2} .
  \end{aligned}
\end{equation}
The numerical coefficients depend on the density power, which indices as \citep{kasen16}
\begin{equation}
  \zeta_{\rho} = \frac{(n - 3) (3 - \delta)}{4 \pi (n - \delta)}, \hspace{1em}
  \zeta_v = \left[ \frac{2 (5 - \delta) (n - 5)}{(n - 3) (3 - \delta)}
  \right]^{1 / 2} .
\end{equation}
For core-collapse SNe, the typical values of the density power indices are $\delta = 1, n = 10$ \citep{chevalier89}. 

With the SN ejecta's expansion, it is expected that a significant fraction of the material in the envelope would enter and be trapped by the gravitational potential of the companion BH. We assume the supernova ejecta is spherically symmetric with $R_{\mathrm{max}}$ as the outer boundary. Once the outer part of the ejecta with $R_{\mathrm{max}}$ reaches the accretion area of the companion BH, the outer part of the SN ejecta with $\rho_{\mathrm{ej}} \propto r^{- n}$ begins to fall into the BH, where $R_{\max}$ at the time $t$ can be expressed by the initial outermost radius ($R_{\mathrm{\max, 0}}$) and the outermost velocity ($v_{\mathrm{ej}, \max}$) as
\begin{equation}
  R_{\mathrm{\max}} = R_{\mathrm{\max, 0}} + v_{\mathrm{ej, \max}} t.
\end{equation}
Similarly, the inner boundary ($R_{\mathrm{\min}}$)
of the ejecta could be defined by the innermost velocity ($v_{\mathrm{ej}, \min}$)
\begin{equation}
  R_{\mathrm{\min}} = R_{\mathrm{\min, 0}} + v_{\mathrm{ej, \min}} t.
\end{equation}

Under super-Eddington condition, the accretion process may have a strong feedback to the SN explosion \citep{Gao2020}. During the accretion process, there are two main feedback mechanisms, including accretion disk radiation and Blandford-Payne outflow \citep{BP}. In this scenario, most of the energy of the Blandford-Znajek (BZ) \citep{BZ} jet is dissipated outside the SN instead of being injected into the SN material \citep{Gao2020}. 

The luminosity produced by the radiation of accretion disk can be expressed as
\begin{equation}
  L_{\mathrm{disk}} = 2 \int_{R_{\mathrm{ms}}}^{R_{\mathrm{out}}} 2 \pi R \sigma
  T_{\mathrm{eff}}^4 d R.
\end{equation}
The effective temperature of the disk can be obtained by considering the evolution of the disk into a multicolored black body \citep{strubbe09}. The disk luminosity due to the super-Eddington accretion can be approximated as $L_{\mathrm{disk}} \sim 0.2 L_{\mathrm{Edd}}\sim 5\times10^{38}{\rm erg~s^{-1}}(M_{\rm BH}/20M_\odot)$. Considering that $L_{\mathrm{disk}}$ is much lower than the power from radioactive element decay ($\sim10^{42} \mathrm{erg~s^{-1}}$), we ignore the contribution of accretion disk radiation in later calculations.

On the other hand, the luminosity generated by the BP outflow could be written as
\begin{equation}
  L_{\mathrm{BP}} (t) = \eta_{\mathrm{BP}} \dot{M}_{\mathrm{tr}} c^2 \left\{
  \begin{array}{ll}
    \left( \frac{t}{t_{\mathrm{tr}}} \right)^{10}, & t_{\mathrm{start}} \leqslant
    t < t_{\mathrm{tr}}\\
    \left( \frac{t}{t_{\mathrm{tr}}} \right), & t_{\mathrm{tr}} \leqslant t
    \leqslant t_{\mathrm{end}}\\
    \left( \frac{t_{\rm end}}{t_{\mathrm{tr}}} \right)e^{-\frac{t-t_{\mathbf{end}}}{t_{\mathbf{end}}}}, &  t
    > t_{\mathrm{end}}
  \end{array} \right.
\end{equation}
where the efficiency factor $\eta_{\mathrm{BP}}$ depends on the BH spin parameter $a$ and $\eta_{\mathrm{BP}} = 0.00876$ for $a=0.5$ is adopted \citep{Gao2020}.
$\dot{M}_{\mathrm{tr}}$ is the falling rate at the characteristic time $t_{\mathrm{tr}}$, which could be approximated as
\begin{equation}
\begin{aligned}
  \dot{M}_{\mathrm{tr}} \simeq &4.1 \times 10^{- 9} M_{\odot}~\text{s}^{- 1}
  \left( \frac{M_{\mathrm{ej}}}{10M_{\odot}} \right)^{5 / 2} \left(
  \frac{M_{\mathrm{BH}}}{20 M_{\odot}} \right)^2 \times\\ 
  &\times\left( \frac{d}{10^{13}
  \mathrm{cm}} \right)^{- 3} \left( \frac{E_{\mathrm{sn}}}{10^{51} \mathrm{erg}}
  \right)^{- 3 / 2}.\label{mff}
  \end{aligned}
\end{equation}
There are three characteristic timescales in the accretion process of the companion BH. $t_{\mathrm{start}}$ is the time for the start of the falling process. $t_{\mathrm{tr}} \sim d / v_{\mathrm{tr}}$ is the time for the falling region reaches the inner part of the  ejecta, which is when the velocity of falling ejecta element $v$ becomes the transition velocity $v_{\mathrm{tr}}$. $t_{\mathrm{end}} \sim d / v_{\mathrm{ej}, \min}$ is taken as the termination timescale of the falling process. After $t_{\mathrm{end}}$, the materials that are marginally bound to BH will continue to move outward on the eccentric orbit and eventually fall back to BH, so the accretion will not stop suddenly but follow an exponential cutoff as $\dot{M}=\dot{M}_{\mathrm{tr}}(t_{\mathrm{end}}/t_{\mathrm{tr}})^{\delta}e^{-(t-t_{\mathrm{end}})/t_{\mathrm{end}}}$.

\begin{figure}[tbph]
\begin{center}
\includegraphics[width=0.49\textwidth,angle=0]{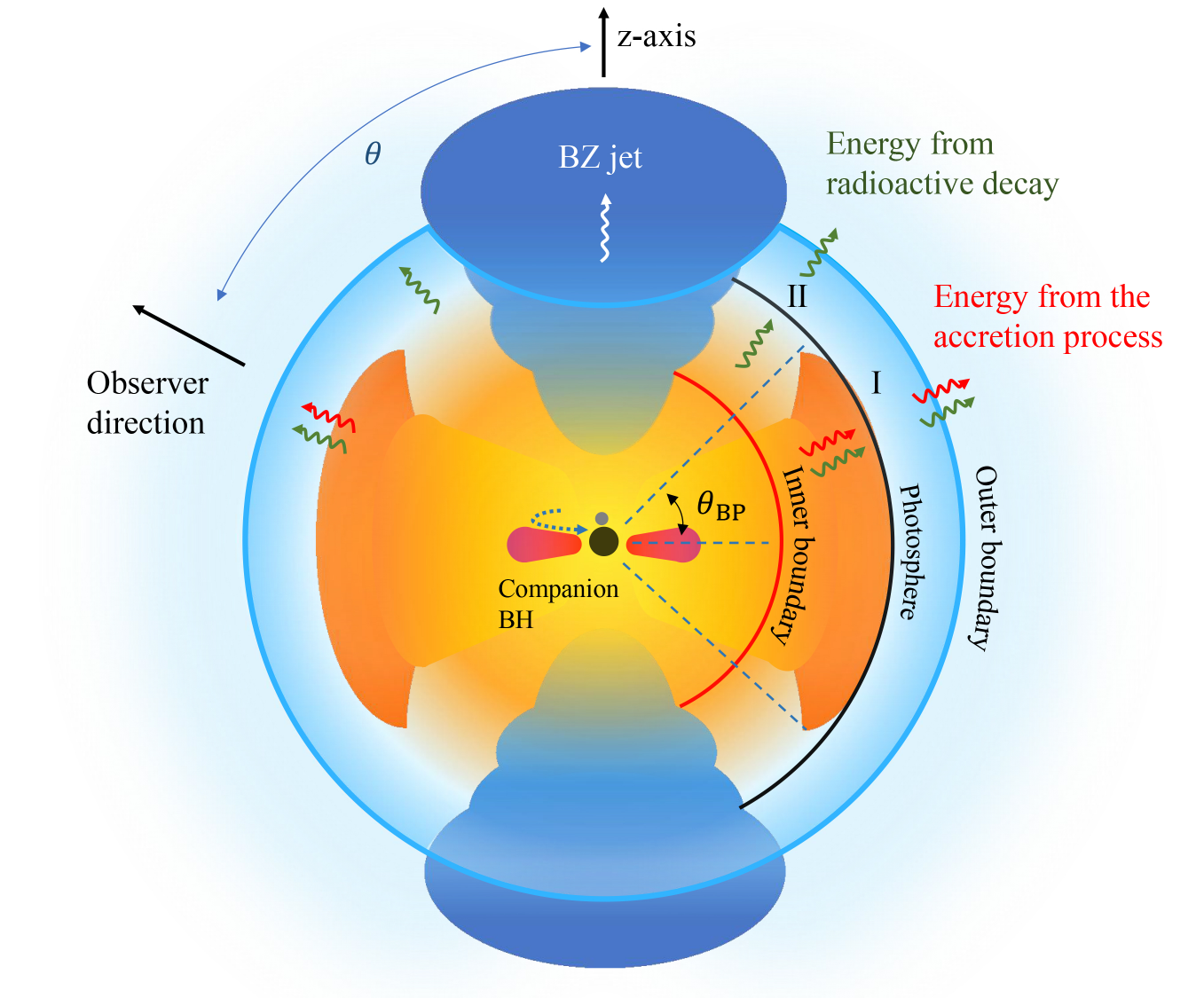}

\end{center}
\caption{An illustration of a special supernova physical model. The companion BH injects energy into the SN ejecta through accretion feedback. The radiation from the BP outflow divides the ejecta photosphere into region \uppercase\expandafter{\romannumeral1} and region \uppercase\expandafter{\romannumeral2} along the angle $\theta_{\rm BP}$ with the equatorial plane, where the energy from region \uppercase\expandafter{\romannumeral1} contains contributions from both the radioactive element decay and the BP outflow.}
\label{fig:Fig_n1}
\end{figure}

The energy from the BP outflow propagates along the magnetic field lines extending away from the accretion disk, ultimately injecting a significant portion of energy into the ejecta along the equatorial direction \citep{BP}. For simplicity and convenience, we assume that the $L_{\mathrm{BP}}$ is mainly concentrated on the photospheric region with a half-opening angle of $\theta_{\rm BP}$ along the equatorial direction (As shown in Figure \ref{fig:Fig_n1}).
Therefore, the radiative region on the photosphere can be divided into Region \uppercase\expandafter{\romannumeral1} ($\theta_{\rm BP} < \theta < \theta_{\rm BP} + \pi/2$) and Region \uppercase\expandafter{\romannumeral2} ($|\pi/2 - \theta| < \theta_{\rm BP}$) based on the polar angle $\theta$. The heating power in region \uppercase\expandafter{\romannumeral1} is provided by the the radioactive decay of $^{56}$Ni and the BP outflow, which can be represented as 
\begin{equation}
  L_{\mathrm{heat},\mathrm{\uppercase\expandafter{\romannumeral1}}} (t) = L_{\mathrm{BP}} (t) + L_{\mathrm{Ni},\mathrm{\uppercase\expandafter{\romannumeral1}}} (t).
\end{equation} 
The heating power in region \uppercase\expandafter{\romannumeral2} is solely contributed by the radioactive decay of $^{56}$Ni, expressed as
\begin{equation}
  L_{\mathrm{heat},\mathrm{\uppercase\expandafter{\romannumeral2}}} (t) = L_{\mathrm{Ni},\mathrm{\uppercase\expandafter{\romannumeral2}}} (t).
\end{equation} 
Here, $L_{\mathrm{Ni},\mathrm{\uppercase\expandafter{\romannumeral1}}/\mathrm{\uppercase\expandafter{\romannumeral2}}}$ represents the heating power from radioactive decay, denoted as 
\begin{equation}
  L_{\mathrm{Ni},\mathrm{\uppercase\expandafter{\romannumeral1}}/\mathrm{\uppercase\expandafter{\romannumeral2}}} (t) = M_{\mathrm{Ni},\mathrm{\uppercase\expandafter{\romannumeral1}}/\mathrm{\uppercase\expandafter{\romannumeral2}}}[(\mathrm{\epsilon_{Ni}}-\mathrm{\epsilon_{Co}})e^{-t/t_{\mathrm{Ni}}}+\mathrm{\epsilon_{Co}}e^{-t/t_{\mathrm{Co}}}],
\end{equation} 
where $\mathrm{\epsilon_{Ni}} = 3.9\times10^{10} \mathrm{erg~g^{-1}~s^{-1}}$ and $\mathrm{\epsilon_{Co}} = 6.8\times10^{9} \mathrm{erg~g^{-1}~s^{-1}}$ are the heating rates of $^{56}$Ni and $^{56}$Co, respectively. $t_{\mathrm{Ni}}$ = 8.8 days and $t_{\mathrm{Co}}$ = 111.3 days are their decay timescales \citep{Khatami2019}.
We simply assume that $^{56}$Ni is uniformly distributed in the ejecta, so the mass of $^{56}$Ni ($M_{\mathrm{Ni},\mathrm{\uppercase\expandafter{\romannumeral1}}}$) contained in region \uppercase\expandafter{\romannumeral1} is
\begin{equation}
  M_{\mathrm{Ni},\mathrm{\uppercase\expandafter{\romannumeral1}}} = \cos(\frac{\pi}{2}-\theta_{\mathrm{BP}}) M_{\mathrm{Ni}},
\end{equation} 
where $M_{\mathrm{Ni}}$ is the mass of all Ni contained in the whole ejecta. The mass of $^{56}$Ni in region $\mathrm{\uppercase\expandafter{\romannumeral2}}$ ($M_{\mathrm{Ni},\mathrm{\uppercase\expandafter{\romannumeral2}}}$) is thus
\begin{equation}
  M_{\mathrm{Ni},\mathrm{\uppercase\expandafter{\romannumeral2}}} = M_{\mathrm{Ni}} - M_{\mathrm{Ni},\mathrm{\uppercase\expandafter{\romannumeral1}}}.
\end{equation} 
The bolometric luminosity for SN radiation from each of the two regions can be estimated as \citep{arnett82}
\begin{equation}
  L_{\mathrm{SN},\mathrm{\uppercase\expandafter{\romannumeral1}}/\mathrm{\uppercase\expandafter{\romannumeral2}}} (t) = e^{- \left( \frac{t^2}{\tau_m^2} \right)} \int_0^t 2
  \frac{t}{\tau_m^2} L_{\mathrm{heat},\mathrm{\uppercase\expandafter{\romannumeral1}}/\mathrm{\uppercase\expandafter{\romannumeral2}}} (t') e^{\left( \frac{t^{\prime
  2}}{\tau_m^2} \right)} d t',
\end{equation}
where $\tau_m$ is the effective diffusion timescale, which reads as
\begin{equation}
  \tau_m = \left( \frac{2 \kappa M_{\mathrm{ej}}}{\beta v c} \right)^{1 / 2}.
\end{equation}
Here $\kappa$ is the opacity of the SN ejecta. $\beta = 13.8$ is a constant for the density distribution of the ejecta  \citep{arnett82}. 
The temperatures of each region can be expressed by assuming that the photosphere is a black body as 
 \begin{gather}
  T_{\mathrm{\uppercase\expandafter{\romannumeral1}}/\mathrm{\uppercase\expandafter{\romannumeral2}}} = \left(\frac{L_{\mathrm{SN},\mathrm{\uppercase\expandafter{\romannumeral1}}/\mathrm{\uppercase\expandafter{\romannumeral2}}}}{\sigma_{\rm SB} A_{\mathrm{\uppercase\expandafter{\romannumeral1}}/\mathrm{\uppercase\expandafter{\romannumeral2}}}}\right)^{1/4} ,
  \label{eq:Temp}
\end{gather}
 where $\sigma_{\rm SB}$ is Steffan-Boltzmann constant. $A_{\mathrm{\uppercase\expandafter{\romannumeral1}}}$ and $A_{\mathrm{\uppercase\expandafter{\romannumeral2}}}$ are the surface areas of region $\mathrm{\uppercase\expandafter{\romannumeral1}}$ and region $\mathrm{\uppercase\expandafter{\romannumeral2}}$ at $R_{\mathrm{ph}}$, respectively. The $R_{\mathrm{ph}}$ satisfies the electron scattering opacity to the outer boundary
\begin{equation}
  \tau(R_{\mathrm{ph}}) = \int_{R_{\mathrm{ph}}}^{R_{\mathrm{max}}} \kappa \rho_{\rm ej}(r,t) d r ,
\end{equation}
 set to $\frac{2}{3}$ as in \cite{arnett82}.

\begin{figure}[tbph]
\begin{center}
\includegraphics[width=0.47\textwidth,angle=0]{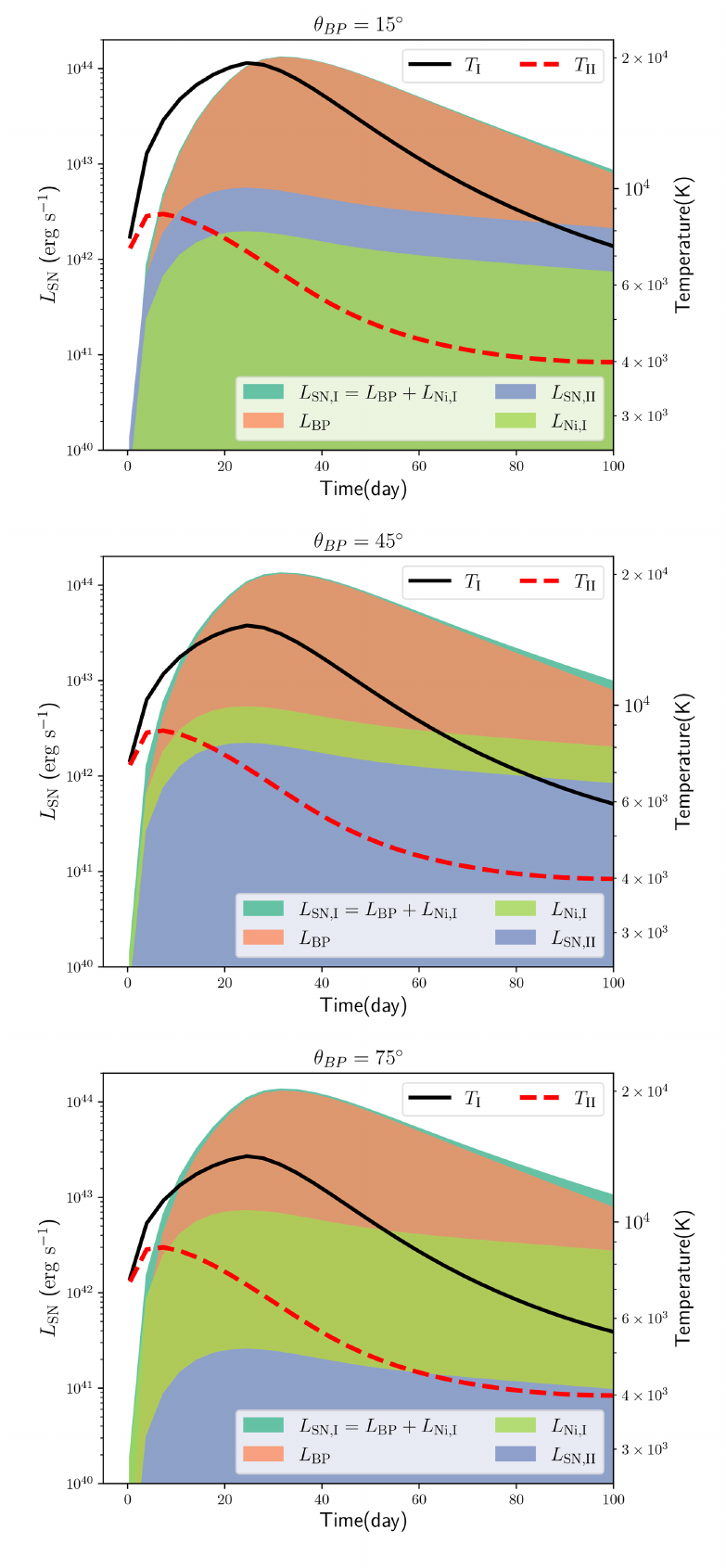}

\end{center}
\caption{The evolution of SN bolometric luminosity and the temperature on the photosphere in Case A. Areas with different color fills show the contribution from different heating mechanisms to the bolometric luminosity. The black and red solid lines represent the temperature of region \uppercase\expandafter{\romannumeral1} and region \uppercase\expandafter{\romannumeral2}, respectively. In our calculation, $d=10^{13}$ cm, $M_{\rm ej}=2 M_{\odot}$, $E_{\rm {sn}}=10^{51}$ erg, $v_{\min} =50 \text{ km s}^{-1}$, $v_{\max} = 0.1 c$, $M_{\rm BH}=20 M_{\odot}$ and $M_{\rm Ni}=0.5 M_{\odot}$ are adopted.}
\label{fig:lightcurve_caseA}
\end{figure}

\begin{figure}[tbph]
\begin{center}
\includegraphics[width=0.47\textwidth,angle=0]{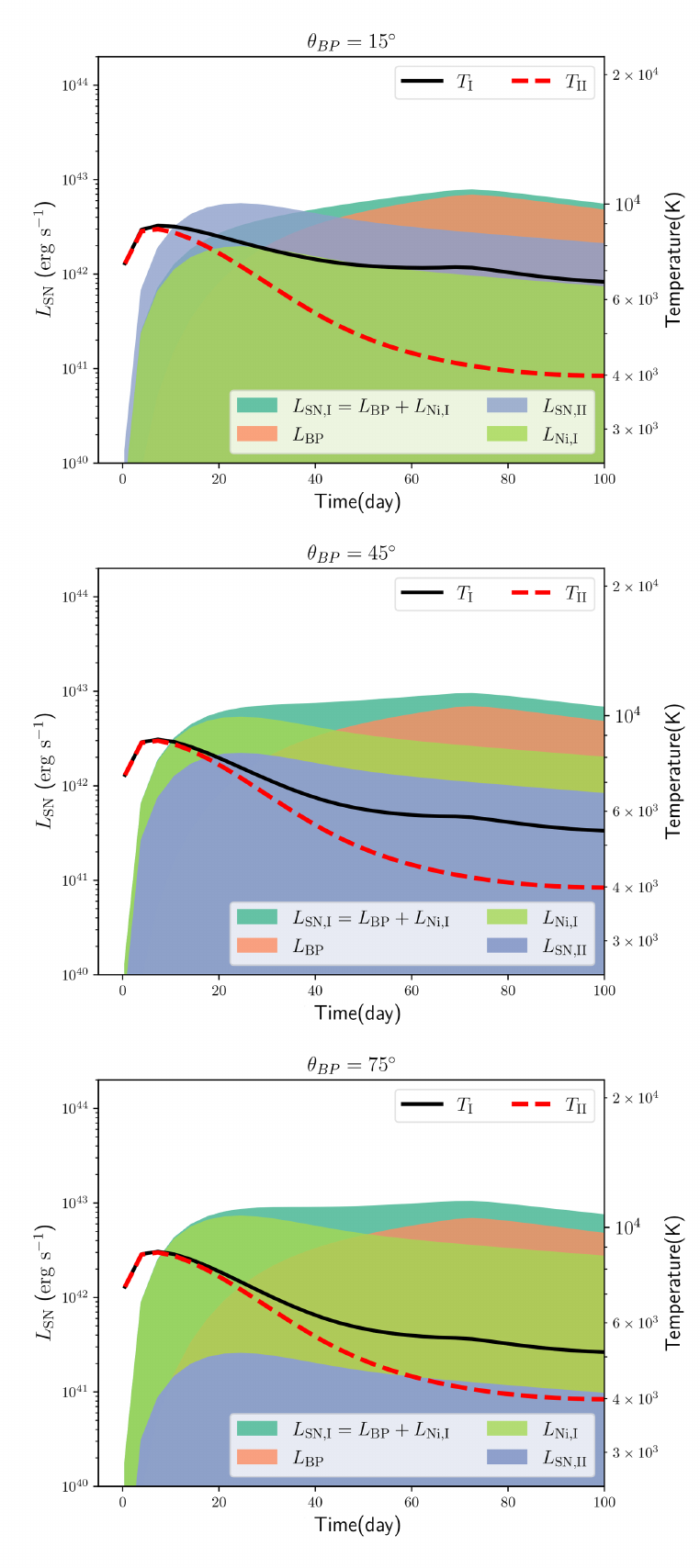}

\end{center}
\caption{
Similar as Figure \ref{fig:lightcurve_caseA} but for Case B. $d = 3\times 10^{13}~{\rm cm}$ is adopted while other model parameters keep unchanged.}
\label{fig:lightcurve_caseB}
\end{figure}

The orbital separation distance $d$ mainly affects the falling rate $\dot{M}_{\mathrm{tr}}$, so that affects the power for BP mechanism.
To facilitate the detailed calculation and analysis, we consider two cases. In Case A (``close-binary" case), we set a small binary separation ($d = 10^{13}{\rm cm}$), so the accretion feedback power is much greater than the radioactive heating power. For this case, Figure \ref{fig:lightcurve_caseA} shows the evolution of the bolometric luminosity and effective temperature on the photosphere for both region I and II. It can be seen that the peak luminosity from region \uppercase\expandafter{\romannumeral1} can sharply reach $10^{44} \rm erg~s^{-1}$,  which is comparable to that of superluminous SNe \citep{galyam2019}. Near the peak, the luminosity from region \uppercase\expandafter{\romannumeral1} is much higher than that from region \uppercase\expandafter{\romannumeral2}. 
The temperature in region \uppercase\expandafter{\romannumeral1} exhibits a decrease as $\theta_{\rm BP}$ increases. This can be attributed to the dispersion of the total injected energy in a larger photoshere region. As the photosphere undergoes rapid expansion, the surface temperature gradually declines over time subsequent to the temperature peak.
In Case B (``faraway-binary" case), we assume a relatively larger binary separation ($d = 3 \times 10^{13}{\rm cm}$), so the accretion feedback power is comparable to the radioactive feedback power. As shown in Figure \ref{fig:lightcurve_caseB}, although the peak luminosity and peak temperature for the two regions are similar, difference might emerge at later phase. Due to the continuous energy injection from the accretion disk via BP mechanism, the temperature evolution of region \uppercase\expandafter{\romannumeral1} would show a plateau characteristic, in which phase the decreasing of SN luminosity from region II is much more significant than that from region I.

\subsection{Polarization} 
The BP mechanism heats the SN ejecta near the equator more efficiently than those near the pole, so that the luminosity distribution on the photosphere is not isotropic, breaking the original spherical symmetry for the SN emission.
Observable polarized signals are then likely to be generated. In order to predict the polarization properties, we develop a 3D Monte Carlo polarization simulation code (MCPSC) where both electron scattering and line scattering are included. The opacity of the SN envelope is mainly caused by electrons' Thompson scattering and bound-bound line transitions \citep{Kasen03}. In this work, we assume the Thompson scattering as the dominant contributor to the continuum polarization in the SNe, which is a special case of the wavelength-dependent Compton scattering at low-frequency limit. The continuum polarization in the optical band is inherently wavelength independent, since Thomson scattering is approximately a grey process at this frequency \citep{Bulla2017}. The line scattering effect is ignored.

Before starting the simulation, we need to do some basic physical configurations. For a beam of radiation emitted from the photosphere, the polarization is described by a useful convention called the Stokes vector  $S = (I, Q, U, V)$, where $I$ gives the total intensity, $Q$ and $U$ measure the degree of linear polarization and \textit{V} measures the degree of circular polarization. Considering that circular polarization has never been observed in SNe, also, in a scattering atmosphere without a magnetic field, the radiative transfer calculations for circular and linear polarization can be decoupled \citep{chandrasekhar1960}, we therefore neglect the $V$ component. The Stokes vector is defined in the plane orthogonal to the direction of radiation propagation $n$ and expressed as 
\begin{equation}
S = 
\left( \begin{array}{c}
I\\ Q\\ U
\end{array} \right) = \left( \begin{array}{c}
I_l  + I_r \\ I_l - I_r \\  I_a - I_b
\end{array} \right) = \left( \begin{array}{c}
\updownarrow + \leftrightarrow \\ \updownarrow - \leftrightarrow  \\ \mathrel{\rotatebox{45}{$\updownarrow$}}-\mathrel{\rotatebox{45}{$\leftrightarrow$}}
\end{array} \right),
\end{equation}
where we introduce two reference axes ($l$ and $r$) to satisfy $n = r \times l$. $l$ lies in the meridian plane (the plane defined by $n$ and the polar axis $z$
) and $r$ is perpendicular to $l$ \citep{Bulla15}.
With this convention, $Q$ is defined as the difference between intensity $I_l$ with electric field oscillating along $l$ and intensity $I_r$ with electric field oscillating along $r$. $U$ is the equivalent difference between the intensities in the directions $a$ and $b$, which are defined by rotating the reference axes $l$ and $s$ counter-clockwisely by $45$ degrees (as viewed looking antiparallel to $n$).
Thus, the dimensionless Stokes vector could be written as
\begin{equation}
s = \frac{S}{I} =\left(\begin{array}{c}
1\\ q\\ u
\end{array} \right) ,
\end{equation}
where $q = Q/I$ and $u = U/I$ are defined as the fractional polarization.
The polarization degree $P$ and the position angle $\chi$ can then be given in terms of the Stokes
Parameters
\begin{equation}
P = \frac{\sqrt{Q^2+U^2}}{I}=\sqrt{q^2+u^2} ~ ,
\end{equation}
\begin{equation}
\label{chi}
\chi = \frac{1}{2} \tan^{-1}{\bigg(\frac{U}{Q}\bigg)}= \frac{1}{2} \tan^{-1}{\bigg(\frac{u}{q}\bigg)} ~ ,
\end{equation}
where $\chi$ is the angle between the electric field orientation and the reference axis ${l}$. 
To conduct the simulation, we follow the electron density distribution assumed in Section 2.1 and assume that the electron density is spherically symmetric. Unpolarized photon packets are launched from random positions on the photosphere with random propagating directions. The propagating direction for a photon packet is chosen randomly with $\mu = \sqrt{z}, z \in (0, 1]$, where $\mu$ is the cosine of the angle between the propagating direction and the radial direction. Then the photon packet propagates in the 3D ejecta and would be scattered by electrons. Here we follow the treatment proposed by previous works \citep{Code95,Mazzali93,Lucy99,Kasen03,Whitney2011,Bulla15} to simulate the electron scattering process.
Scattering changes the Stokes vector for a photon packet, through a converting matrix (scattering matrix).
The probability for a photon packet to be scattered to a certain direction is given by the probability distribution of scattering angles depending on the Stokes vector. When the photon packet reaches the outer boundary of the ejecta, record its propagating angle at that time. Finally, it will be collected by an observer with the corresponding observing angle.
Under the spherical symmetry condition, we divide the line of sight into 16 bins along the polar angle, and it is not necessary to bin in longitude due to the symmetry. More details for the simulating procedure and code are given in Appendix.

\section{Results}

\begin{figure*}
\centering
\includegraphics[width=1.0\textwidth]{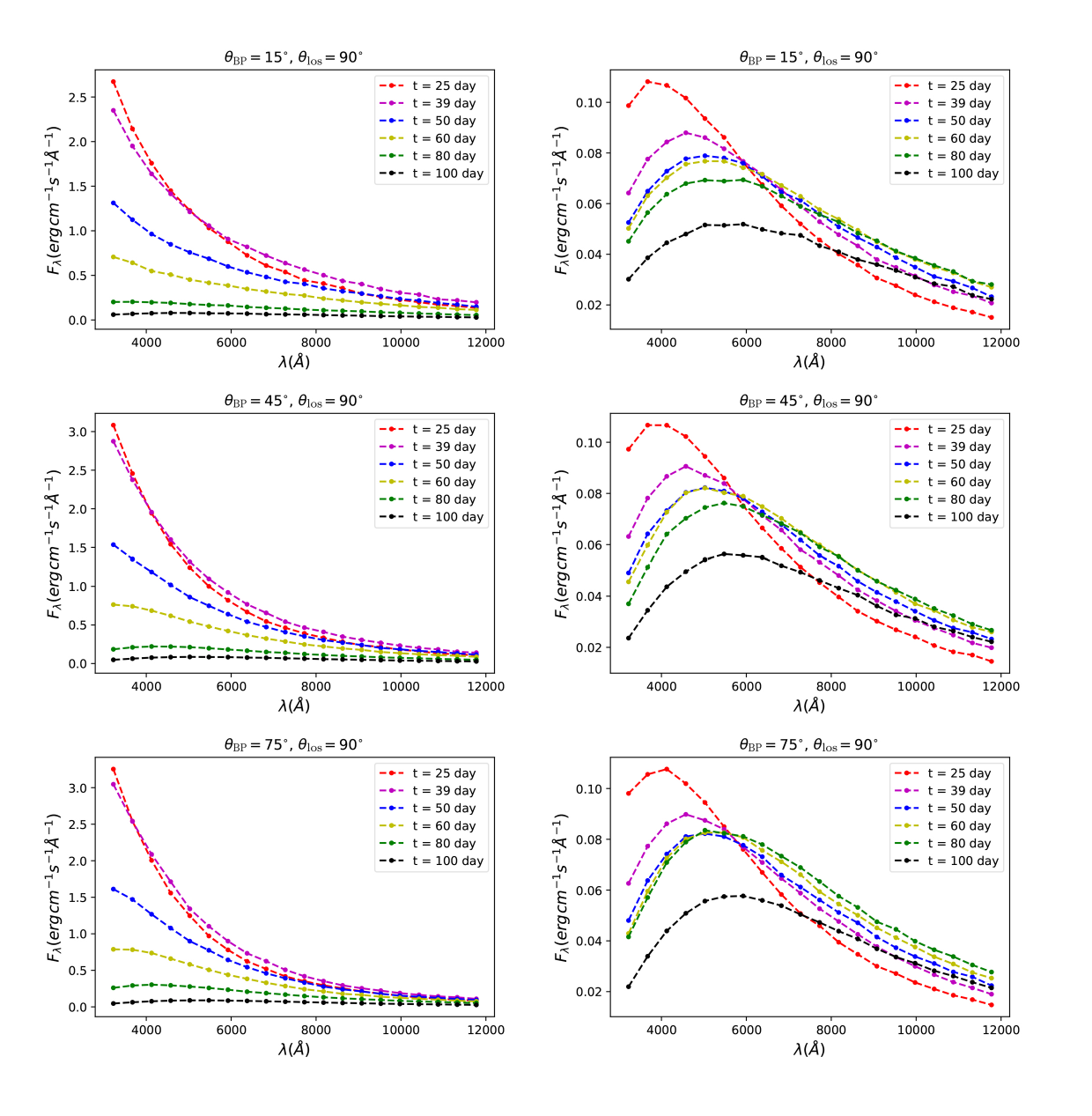}
\caption{The emergent spectra of Case A (left panel) and Case B (right panel) in the line of sight direction $\theta_{\rm los} = 90^{\circ}$, where a distance of 10 pc is adopted. The panels from top to bottom show three angular configurations for area \uppercase\expandafter{\romannumeral1}: $\theta_{\rm BP} = 15^{\circ}$,$\theta_{\rm BP} = 45^{\circ}$ and $\theta_{\rm BP} = 75^{\circ}$ respectively. Each panel shows the emergent spectra on days 25, 39, 50, 60, 80 and 100 after the supernova explosion with different coloured dot-dashed lines.
}
\label{fig:fig01}
\end{figure*}

\begin{figure*}
\centering
\includegraphics[width=1.0\textwidth]{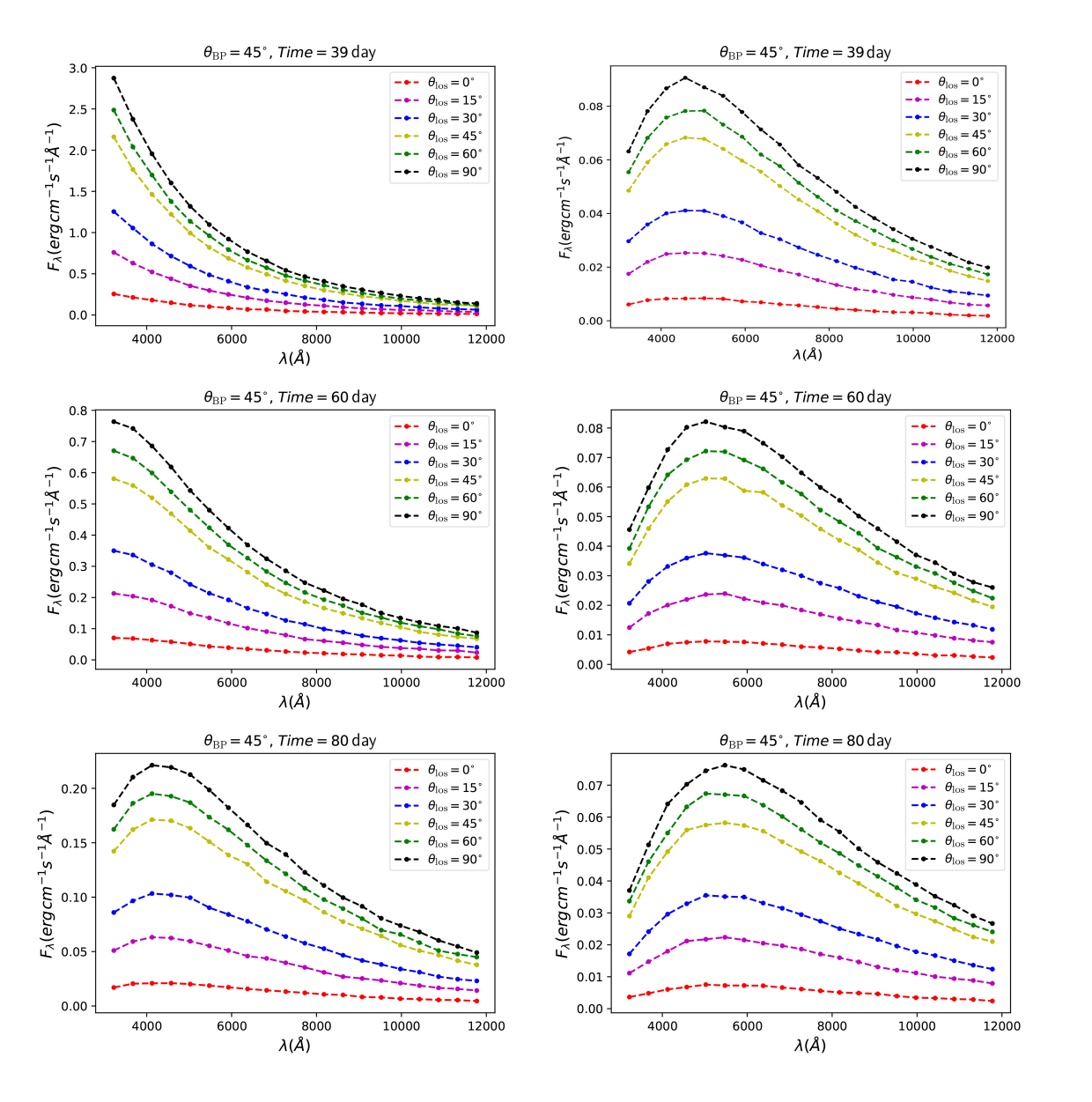}
\caption{
The emergent spectra of Case A (left panel) and Case B (right panel) with the same angular configuration ($\theta_{\rm BP} = 45^{\circ}$) for area \uppercase\expandafter{\romannumeral1}, where a distance of 10 pc is adopted. The panels from top to bottom show the emergent spectra on days 39, 60 and 80 after the supernova explosion, respectively. Each panel shows the emergent spectrum in the line of sight direction $\theta_{\rm los}$ for $0^{\circ}$,$15^{\circ}$,$30^{\circ}$,$45^{\circ}$,$60^{\circ}$ and $90^{\circ}$ respectively, using different coloured dotted dashed lines. 
}
\label{fig02}
\end{figure*}

We performed radiative transfer simulations using MCPSC code and collected photons along the line of sight (LOS) direction to obtain the spectral and polarizing features of the SNe. Figure \ref{fig:fig01} shows the emergent spectra for Case A and Case B at a distance of 10 pc from the source, when the SN is observed with the viewing angle $\theta_{\rm los}$ $=90^{\circ}$. In Case A, the flux density only slightly increases with a greater $\theta_{\rm BP}$ in the optical band, while in Case B, the increasing of flux density with a greater $\theta_{\rm BP}$ is obvious at $t \sim 50 - 80$ days (plateau phase). The emergent spectra observed along different LOS directions for a given $\theta_{\rm BP}$ at different observing time are shown in Figure 5. Overall, under the same $\theta_{\rm BP}$ and with same observing angle $\theta_{\rm los}$, the spectrum of the SN in Case B, due to its relatively lower effective temperature, would be redder than that in Case A.

Then, we discuss the continuum polarization property for the SNe in Case A and Case B separately.
The radioactive-decay process uniformly heats the ejecta in regions \uppercase\expandafter{\romannumeral1}  and \uppercase\expandafter{\romannumeral2}, which itself might power an isotropic thermal radiation from the photosphere.
The BP mechanism could provide additional internal energy to region \uppercase\expandafter{\romannumeral1} , making thermal radiation from region \uppercase\expandafter{\romannumeral1} significantly brighter than that from region \uppercase\expandafter{\romannumeral2}, breaking the isotropy. Considering that a photon packet experiences one or several times of scattering, and is finally collected by the observer, its polarization property might have changed. Since the SN envelope is assumed to be spherically symmetric, for an observer with arbitrary viewing angle, there would always be
two directions, from which the two photon packets should have opposite polarization properties and finally can be cancelled out by each other. However, because more photons will be released from the photosphere in region \uppercase\expandafter{\romannumeral1} than region \uppercase\expandafter{\romannumeral2}, usually the polarization cannot be totally cancelled out.

\begin{figure}[tbph]
\begin{center}
\includegraphics[width=0.49\textwidth,angle=0]{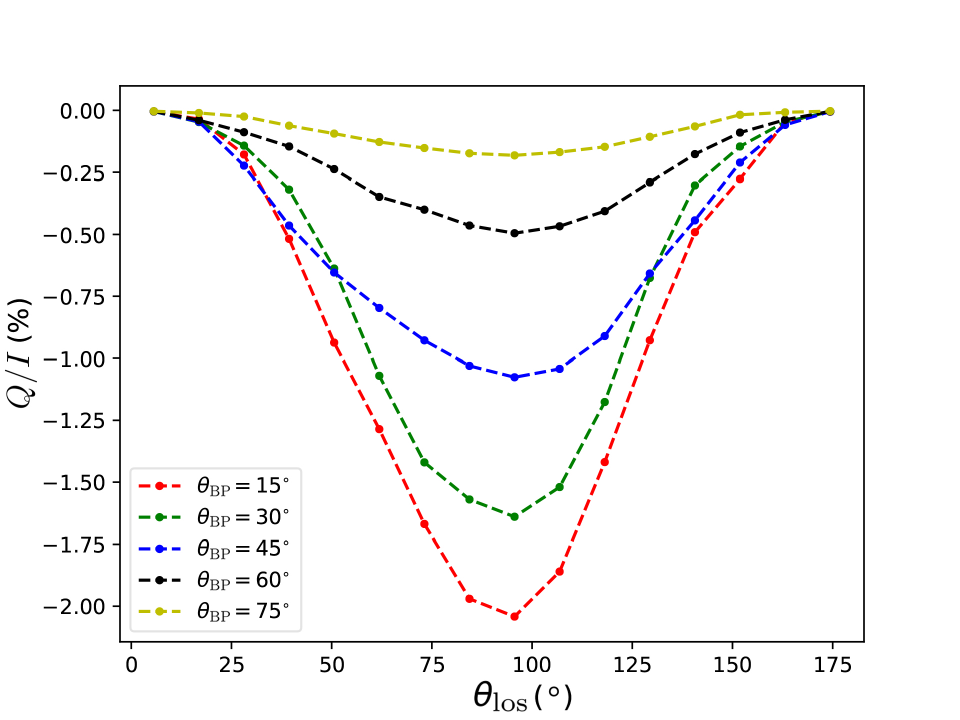}

\end{center}
\caption{The continuum polarization degree observed on day 39 after the SN explosion in Case A at different viewing angles $\theta_{\rm los}$, where the color of the curve represents the continuum polarization degree for different $\theta_{\rm BP}$ of region \uppercase\expandafter{\romannumeral1}. The Stokes vector U is 0 due to symmetry.}
\label{fig:polarization degree 1}
\end{figure}

In case A, near the peak of the SN, the contribution to the total luminosity from region \uppercase\expandafter{\romannumeral1}, is much higher than that from region \uppercase\expandafter{\romannumeral2}. Their significant difference in luminosity serves as the main reason for the emergence of continuum polarization, which highly depends on the the viewing angle ($\theta_{\rm los}$) and the opening angle of region \uppercase\expandafter{\romannumeral1} ($\theta_{\rm BP}$). Denote the polarization degree as $P = Q/I$, where the Stokes vector $U$ is 0 due to symmetry.
We calculate the continuum polarization properties produced at 39 days after the SN explosion, as shown in Figure \ref{fig:polarization degree 1}. Here we choose $\theta_{\rm BP} = (15^{\circ}, 30^{\circ}, 45^{\circ}, 60^{\circ}, 75^{\circ})$ and a series of $\theta_{\rm los}$ from $0^{\circ}$ to $180^{\circ}$.
This case allows to produce a significantly polarized signal in some conditions, with a maximum continuous polarization degree ($P_{\rm max}$) as about $\sim 2\%$ when $\theta_{\rm BP} = 15^{\circ}$ and $\theta_{\rm los} = 90^{\circ}$. 
$P$ decreases as $\theta_{\rm BP}$ increases and approaches a maximum when the luminosity is mainly distributed near the equator like $\theta_{\rm BP} = 15^{\circ}$. The continuum polarization tends to be approximately zero at $\theta_{\rm BP} \geq 75^{\circ}$ where the luminosity becomes quasi-isotropic.
On the other hand, with a fixed $\theta_{\rm BP}$, the polarization degree peaks at $\theta_{\rm los} = 90^{\circ}$ (near the equator). This is because the projection area of region \uppercase\expandafter{\romannumeral1} in the LOS direction is the largest at around $\theta = 90^{\circ}$, where the contribution from the high luminosity part with significant polarization to the finally observed SN signal reaches the maximum. As the LOS gets closer to the polar direction, the polarization degree gradually decreases and eventually falls to $0 \%$.

\begin{figure}[tbph]
\begin{center}
\includegraphics[width=0.49\textwidth,angle=0]{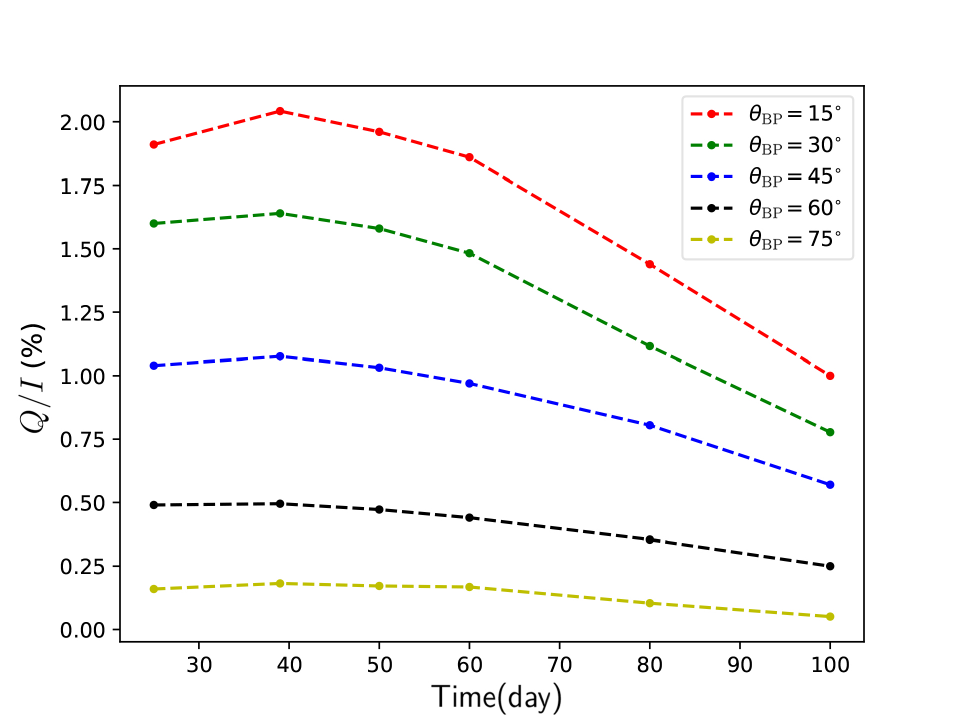}

\end{center}
\caption{The evolution of polarization degree with time after the supernova explosion in Case A, where the color of the curve represents the continuum polarization degree for the different $\theta_{\rm BP}$ at $\theta_{\rm los} = 90^{\circ}$.}
\label{fig:polarization curve1}
\end{figure}

The evolution of the polarization degree with time is shown in Figure \ref{fig:polarization curve1}. As the BP mechanism gradually dominates the SN emission,
$P$ also increases with time and reaches a maximum value at the light curve's peak, after which the $P$ decreases with time since the BP power decays faster than the radioactive power. We calculate the polarization degrees at 25, 39, 50, 60, 80 and 100 days after the SN explosion.
For comparison purposes, we also consider the conditions with $\theta_{\rm BP} = (15^{\circ}, 30^{\circ}, 45^{\circ}, 60^{\circ}, 75^{\circ})$.
Here we fix the viewing angle as $\theta_{\rm los} = 90^{\circ}$ to make the polarization as recognizable as possible. Even so, for large $\theta_{\rm BP}$ (i.e. $\theta_{\rm BP} \ge 80^{\circ}$), we find it is difficult to identify the evolution of polarization degree with time, because $P$ is too close to $0\%$ after the peak of the SN light curve. The cases with different $\theta_{\rm BP}$ have the similar evolving trend.

\begin{figure}[tbph]
\begin{center}
\includegraphics[width=0.49\textwidth,angle=0]{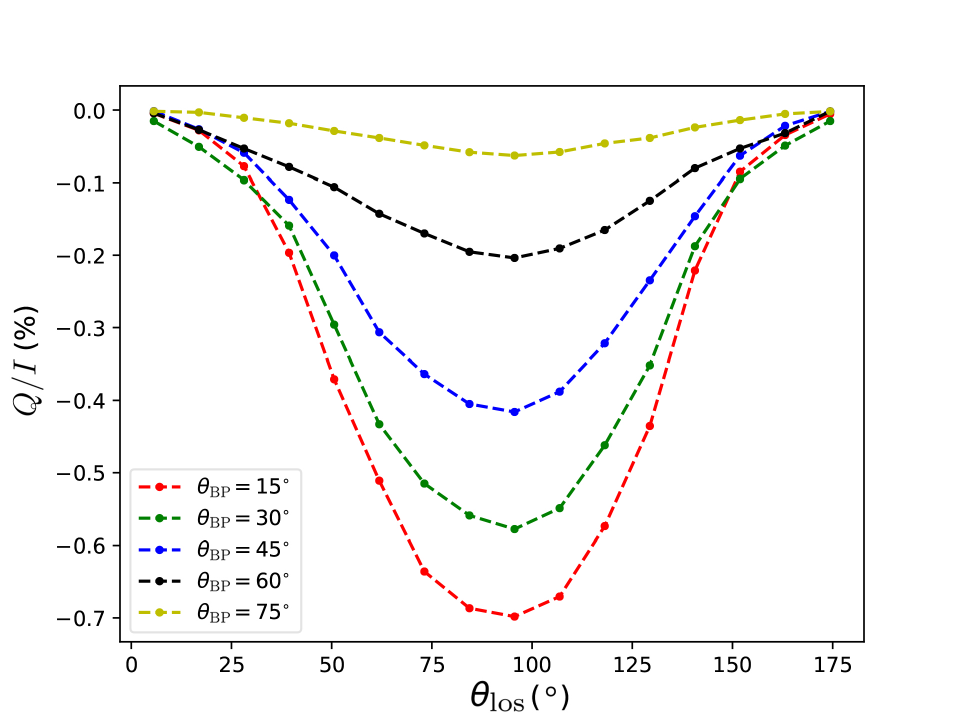}

\end{center}
\caption{The continuum polarization degree observed on day 39 after the SN explosion in Case B at different viewing angles $\theta_{\rm los}$, where the color of the curve represents the continuum polarization degree for different $\theta_{\rm BP}$ of region \uppercase\expandafter{\romannumeral1}. The Stokes vector U is 0 due to symmetry.}
\label{fig:polarization degree2}
\end{figure}

\begin{figure}[tbph]
\begin{center}
\includegraphics[width=0.49\textwidth,angle=0]{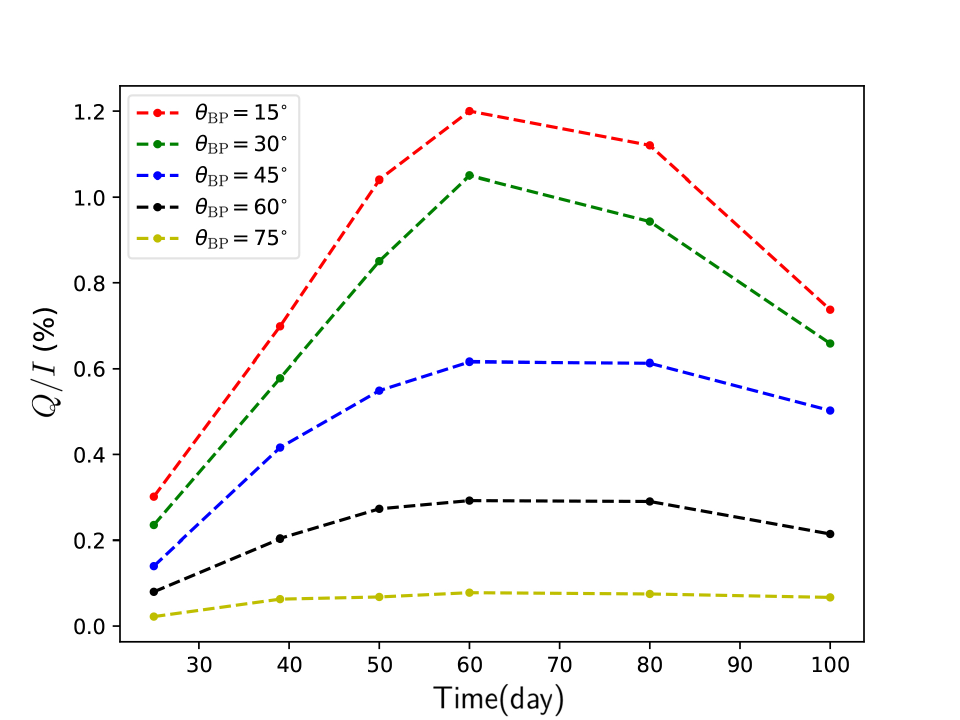}

\end{center}
\caption{The evolution of polarization degree with time after the supernova explosion in Case A, where the color of the curve represents the continuum polarization degree for the different $\theta_{\rm BP}$ at $\theta_{\rm los} = 90^{\circ}$.}
\label{fig:polarization curve2}
\end{figure}

In Case B, the BP effect only becomes dominant at at a later time when the radioactive power has significantly decayed. Therefore, the polarization degree of the SN signal should also be higher in the plateau phase in Figure \ref{fig:lightcurve_caseA}, when the non-isotropy of the luminosity arises. Again, we firstly study the continuum polarization property for the signal at a certain time with different $\theta_{\rm BP}$ and $\theta_{\rm los}$. Then, study the evolution of the polarization degree with time.
Figure \ref{fig:polarization degree2} shows the $P$ value's dependence on $\theta_{\rm BP}$ and $\theta_{\rm los}$, at $39$ days after the SN explosion. Similar to that in Case A, we calculated the $P$ values under different $\theta_{\rm BP}$ and $\theta_{\rm los}$ values. Generally, $P$ values tend to be smaller than that in Case A, because the luminosity in the two regions have the same order of magnitude.
For a given LOS direction, in the absence of very significant brightness differences between the two regions on the photosphere, the polarization level depends mainly on the geometric distribution of the polarization vector in the projection plane, which is determined by $\theta_{\rm BP}$. At $\theta_{\rm BP}=15^{\circ}$, $P_{\rm max}$ can reach about $0.7\%$, which is smaller than the $P_{\rm max} \sim 2 \%$ in Case A, but still may be identified in observation. 
The dependency of $P$ on factors $\theta_{\rm los}$ and $\theta_{\rm BP}$ is evident, with higher $P$ values predominantly observed for lower values of $\theta_{\rm BP}$ and at observation angles near the equator (approximately $\theta_{\rm los} \pm 90$).
Figure \ref{fig:polarization curve2} reveals the time evolution of P, showing how it behaves as a time-delayed feature after the supernova explosion.
At the beginning of the explosion, $P$ remains at a low level because the BP mechanism had not yet produced a significant feedback effect. As the light curve enters the plateau phase, polarization starts to be generated and reaches its peak with the enhancement of the BP feedback and the weakening of the radioactive decay heating. After the peak, both of the powers decline and $P$ starts to decrease as a result.

\begin{figure}[tbph]
\begin{center}
\includegraphics[width=0.49\textwidth,angle=0]{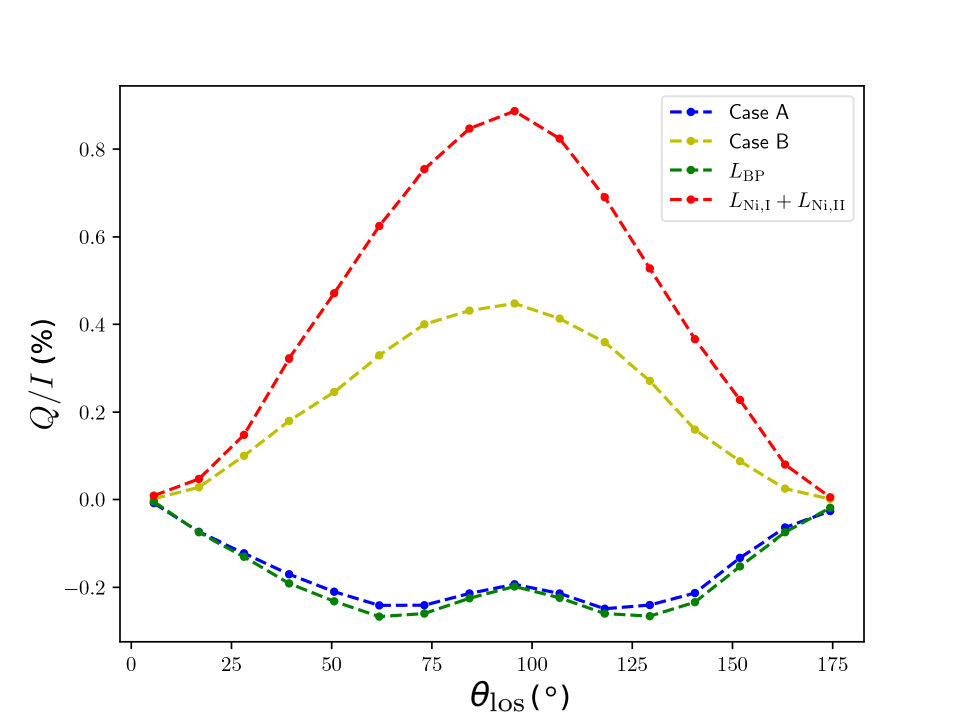}

\end{center}
\caption{The continuum polarization degree observed at 39 days after the SN explosion in oblate elliptic geometry at different $\theta_{\rm los}$. The blue and yellow lines indicate the polarization degree in the Case A and Case B cases, respectively. The red line is the polarization degree produced by the non-spherical ejecta geometry powered by the decay of radioactive elements only. The green line is the polarization degree from the BP mechanism only.}
\label{fig:oblate}
\end{figure}

\begin{figure}[tbph]
\begin{center}
\includegraphics[width=0.49\textwidth,angle=0]{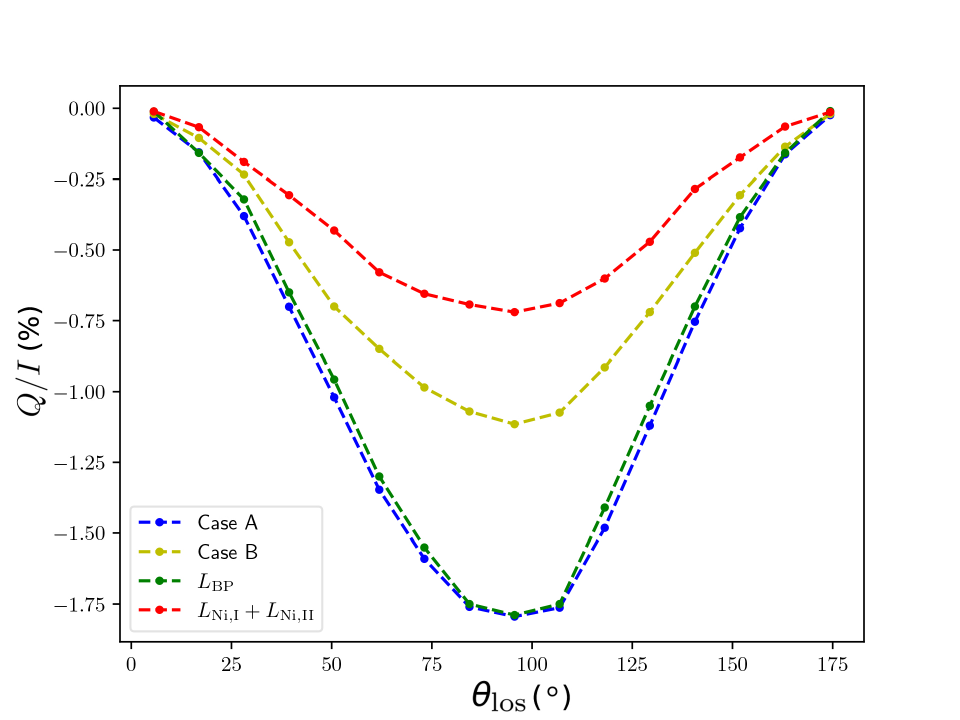}

\end{center}
\caption{Similar as Figure 10 but for prolate elliptic geometry.}
\label{fig:prolate}
\end{figure}

Further, we test the impact of non-spherical ejecta structures on observed polarization levels. In principle, a non-spherically symmetric ejecta could generate an additional polarization signal due to geometric effects.
In some cases, the ejecta along the polar direction may experience faster expansion due to the strong jet-energy injection \citep{Soker22}. There is another possibility that, when the BP mechanism is the main contributor to the energy in region near equator, the ejecta in this region may expand faster than that along the polar direction. Therefore, the envelope could either have a prolate ellipsoidal or oblate ellipsoidal shape, whose geometry can be generally described as
\begin{equation}
\frac{x^2}{a^2}+\frac{y^2}{b^2}+\frac{z^2}{c^2}=1~,~~a=b~~,
\end{equation}
where the prolate elliptic and oblate elliptic geometries can be defined respectively in cylindrical coordinates as
\begin{alignat}{2}
\frac{r^{2}}{A_{1}^{2}}+z^{2}&=R_{\rm max}^{2}, ~~A_{1}=a/c&\quad&(\rm prolate)\\
r^{2}+\frac{z^{2}}{A_{2}^{2}}&=R_{\rm max}^{2}, ~~A_{2}=c/a&&(\rm oblate).
\end{alignat}
We adopt the axis ratio as $A_{1}=A_{2}=0.8$ and calculate the continuum polarization at 39 days after the supernova explosion under the conditions of Case A and Case B, respectively. Figure \ref{fig:oblate} illustrates the results of $P$ for the oblate ellipsoid case at different viewing angles $\theta_{\rm los}$ with $\theta_{\rm BP} = 45^{\circ}$. To facilitate a comprehensive comparison, we evaluate the polarization signal solely generated by the distorted geometry for the ejecta, under the energy input from radioactive element decay (red line in Figure \ref{fig:oblate}), and the polarization signal fully resulting from the BP mechanism in the region \uppercase\expandafter{\romannumeral1} (green line in Figure \ref{fig:oblate}). 
The polarization signal from the BP mechanism appears as a double-humped profile over the viewing angle due to the effect of the ejecta geometry distortion (reaching a maximum value of $P$ at around $90^{\circ}\pm30^{\circ}$).
In Case A, the observed polarization is dominated by the BP mechanism, despite the fact that non-spherical ejecta could produce highly polarized signals with opposite signs. However, in Case B, the polarization signal mainly comes from the geometric effect because the luminosity from the BP mechanism is smaller than that from the decay of radioactive elements. The presence of the BP mechanism could further weaken the polarization level from the geometry distortion.
The polarization from the prolate ellipsoid is significantly different from that from the oblate ellipsoid (as shown in Figure \ref{fig:prolate}). The prolate ellipsoidal ejecta geometry could have the same sign as the polarization signal generated by the BP mechanism, which implies
that these two effects can be mutually reinforcing 
(Case B in Figure \ref{fig:prolate}). In Case A, the polarization is completely dominated by the BP mechanism. In general, the observed polarization levels depend mainly on the total contribution of the two polarization-generating channels, the ejecta's geometric distortion and the BP mechanism. 
Although the polarization could be generated only by the ejecta's geometric distortion, BP mechanism would bring more variability.

\section{Conclusion and Discussion}

In this study, we propose that a SN that is fed by a companion compact star would exhibit non-uniform distribution of luminosity on its photosphere. This can be attributed to the BP mechanism, which serves as a crucial mechanism for accretion feedback, leading to heating of the SN ejecta along the equatorial direction within a confined opening angle ($\theta_{\rm BP}$). We find that the non-uniform distribution of luminosity may serve as a promising novel source of polarization. This is a distinct phenomenon from the well-established factors, such as variations from spherically symmetric photospheres \citep{Shapiro82,Hoflich91,Hoflich96,Dessart11,Bulla15}, the presence of obstructing materials at the photosphere \citep{Kasen03,Hole10,Tanaka2017}, and the existence of off-center radiation sources within the ejecta\citep{Hoflich95}.

The characteristics of the polarization signal for a engine-fed SN depends mainly on the properties of the binary system, the half-opening angle of the radiative region, and the viewing angle. 
Firstly, the distance between the binary orbits actively influences the polarization degree level and time-evolving characteristic by impacting both the accretion rate and the accretion start time of the companion BH.
When the binary is close, the heating power produced by the BP mechanism is much higher than that from the radioactive decay. Consequently, observers in the equatorial direction could witness a significant continuum polarization level in the early days of the supernova explosion. However, as the distance between the binary stars increases, the polarization level gradually diminishes, and the peak polarization occurs at later times.
Moreover, for a given line of sight, smaller radiation cone opening angles, which represent the region where accretion feedback acts upon the ejecta, are more likely to generate higher levels of polarization. This is because the energy from the accretion feedback is more concentrated into the projectile, resulting in more significant differences in the photospheric surface luminosity.
Lastly, the viewing angle adds another layer of complexity in determining the level of polarization. 
Distinct polarization features are observed from various LOS directions, with the maximum polarization occurring at $\theta_{\rm los} = 90^{\circ}$ and the minimum polarization occurring at $\theta_{\rm los} = 0^{\circ}$.
While parameter degeneracy is more likely to occur at low $P$ values, systems with larger $P$ values will exhibit a more prominent angular dependence, indicating the polarimetry as a potential tool to probe angle-dependent information \citep{Bulla2022}.

In addition, we also consider that the engine-fed SN ejecta may also be susceptible to distortion, resulting in a non-spherical shape. In this scenario, the extent of observed polarization is reliant on the interplay between the two channels responsible for generating polarization, as each channel vies for dominance in determining the overall polarization. Specifically, in the context of an oblate ellipsoidal geometry, the polarization direction caused by geometric distortion is antipodal to that produced by non-uniform luminosity. As a result, the net polarization degree is subject to cancellation between these two channels. On the other hand, polarization generated by geometric distortion in the prolate ellipsoidal geometry aligns with that generated by non-uniform luminosity, amplifying the overall polarization degree.

We need to point out that there are still some uncertainties in the results of this work, which would require future complex numerical simulations to fully address. For instance, although there may be cases where a mild transition between two radiation zones will not have a clear demarcation line, it remains valid to consider these cases by assigning specific $\theta_{\rm BP}$ values and $L_{\mathrm{heat},\mathrm{\uppercase\expandafter{\romannumeral1}}}$/$L_{\mathrm{heat},\mathrm{\uppercase\expandafter{\romannumeral2}}}$. On the other hand, the distribution of $^{56}$Ni within the ejecta may not be uniform in practical situations. This non-uniform distribution can have a direct impact on both the magnitude and spatial distribution of luminosity, consequently affecting the observed polarization. Finally, if there is an intermittent black hole accretion process, or a slower rate of accretion as the black hole passes through the innermost ejecta, the lightcurve would be more complex \citep{Gao2020}. The polarization, which is dependent on luminosity variations, would then oscillate over time or have a shallower feature of magnitude at peaks or plateaus.

At present, significant progress has been made in the observation of supernova polarization \citep{Brown2016,Inserra2016,Cikota2018,Lee2019,Lee2020,Leloudas2015,Leloudas2017,Maund2019,Maund2020,Maund2021,Saito2020}. In future surveys, companion-fed SNe could be effectively searched for through the combination of their unique photometric behavior and special polarization properties. Comparing the event rate density of these special SN signals with the event rate density of LIGO-Virgo detected BH–NS/BH systems could help to distinguish the BH–NS/BH formation channel.

\acknowledgments
We are grateful to P. H{\"o}flich for sharing his code and providing some helpful suggestions. This work is supported by the National Natural Science Foundation of China (Projects 12021003,11833003), the National SKA Program of China (2022SKA0130100,2020SKA0120300), and the National Key R\&D Program of China (2021YFA0718500).

\bibliographystyle{apj} 

\begin{thebibliography}{}
\expandafter\ifx\csname natexlab\endcsname\relax\def\natexlab#1{#1}\fi

\bibitem[Abbott(2016)]{Abbott2016}Abbott, B. P., et al. 2016, \prl, 116, 061102
\bibitem[Abbott(2017a)]{Abbott2017a}Abbott, B., et al. 2017a, \prl, 119, 161101
\bibitem[Abbott(2017b)]{Abbott2017b}Abbott, B., et al. 2017b, \apjl, 848, L12
\bibitem[Abbott(2020)]{GW190425}Abbott, B., Abbott, R., Abbott, T.D., et al.\ 2020a, \apjl, 892, L3

\bibitem[Abbott(2021a)]{Abbott2021a}Abbott, B., et al. 2021a, Phys. Rev. X, 11, 021053
\bibitem[Abbott(2021b)]{Abbott2021b}Abbott, B., et al. 2021b, arXiv:2111.03606

\bibitem[Akutsu(2021)]{Akutsu2021}Akutsu, T., Ando, M., Arai, K., et al. 2021, PTEP, 2021, 05A101
A

\bibitem[Anand et al.(2020)]{Anand20} Anand, S., Coughlin, M.~W., Kasliwal, M.~M., et al.\ 2020, Nature Astronomy, doi:10.1038/s41550-020-1183-3
\bibitem[Arnett(1982)]{arnett82} Arnett, W.~D.\ 1982, \apj, 253, 785

\bibitem[Bardeen et al.(1972)]{bardeen72} Bardeen, J.~M., Press, W.~H., \& Teukolsky, S.~A.\ 1972, \apj, 178, 347
\bibitem[Belczynski et al.(2016)]{belczynski16} Belczynski, K., Repetto, S., Holz, D.~E., et al.\ 2016, \apj, 819, 108
\bibitem[Brown et al.(2016)]{Brown2016}Brown, P. J., et al.\ 2016, \apj, 828, 3

\bibitem[{Blandford \& Znajek}(1977)]{BZ}Blandford, R. D., \& Znajek, R. L., 1977, MNRAS, 179, 433
\bibitem[{Blandford \& Payne}(1982)]{BP}Blandford, R. D., \& Payne, D. G., 1982, MNRAS, 199, 883
\bibitem[Bulla et al.(2015)]{Bulla15} Bulla, M., Sim, S. A., \& Kromer, M. 2015, \mnras, 450, 967
\bibitem[Bulla(2017)]{Bulla2017} Bulla, M.\ 2017, PhD thesis, Astrophysics Research Centre, School of Mathematics and Physics, Queen’s University Belfast, Belfast BT7 1NN, UK

\bibitem[Bulla(2022)]{Bulla2022}Bulla, M., Coughlin, M. W., Dhawan, S., \& Dietrich, T. 2022, Universe, 8, 289
\bibitem[Castor(1970)]{castor70} Castor, J.~I.\ 1970, \mnras, 149, 111

\bibitem[Chandrasekhar(1960)]{chandrasekhar1960} Chandrasekhar, S.\ 1960, Radiative Transfer. Dover Press, New York
\bibitem[Chevalier \& Soker(1989)]{chevalier89} Chevalier, R.~A. \& Soker, N.\ 1989, \apj, 341, 867

\bibitem[Cikota et al(2018)]{Cikota2018}Cikota, A., et al.\ 2018, \mnras, 479, 4984

\bibitem[Code \& Whitney(1995)]{Code95} Code A. D., Whitney B. A.\ 1995, \apj, 441, 400

\bibitem[Daniel(1980)]{Daniel1980}Daniel, J. Y. 1980, \aap, 86, 198

\bibitem[Dessart \& Hillier(2011)]{Dessart11} Dessart, L., \& Hillier, D. J.\ 2011, \mnras, 415, 3497

\bibitem[Gal-Yam(2019)]{galyam2019} Gal-Yam, A.\ 2019, \araa, 57, 305

\bibitem[Gao et al.(2020)]{Gao2020} Gao, H., Liu, L.-D., Lei, W.-H., et al.\ 2020, \apjl, 902, L37

\bibitem[Hillier(1991)]{Hillier91}Hillier, D. J.\ 1991, \aap, 247, 455
\bibitem[H{\"o}flich(1991)]{Hoflich91} H{\"o}flich, P.\ 1991, \aap, 246, 481

\bibitem[H{\"o}flich et al.(1996)]{Hoflich96} H{\"o}flich, P., Wheeler, J. C., Hines, D. C., et al.\ 1996, \apj, 459, 307

\bibitem[H{\"o}flich et al.(1995)]{Hoflich95} H{\"o}flich, P.\ 1995, \apj, 440, 821

\bibitem[Hole(2010)]{Hole10} Hole, K. T., Kasen, D., \& Nordsieck, K. H.\ 2010, \apj, 720, 1500
\bibitem[Inserra et al.(2016)]{Inserra2016}Inserra, C., Bulla, M., Sim, S. A., \& Smartt, S. J. 2016,
ApJ, 831, 79
\bibitem[Jeffery(1989)]{Jeffery1989} Jeffery, D. J.\ 1989, ApJS, 71, 951

\bibitem[Jeffery \& Branch(1990)]{Jeffery1990} Jeffery, D. J., \& Branch, D.\ 1990, in Supernovae, ed. J. C. Wheeler, T. Piran, \& S. Weinberg (Singapore: World Scientific), 149

\bibitem[Khatami(2019)]{Khatami2019} Khatami, D. K., \& Kasen, D. N.\ 2019, \apj, 878, 56

\bibitem[Kasen(2003)]{Kasen03} Kasen, D., et al.\ 2003, \apj, 593, 788
\bibitem[Kasen \& Bildsten(2010)]{kasen10} Kasen, D., \& Bildsten, L.\ 2010, \apj, 717, 245

\bibitem[Kasen et al.(2016)]{kasen16} Kasen, D., Metzger, B.~D., \& Bildsten, L.\ 2016, \apj, 821, 36

\bibitem[Lee(2019)]{Lee2019}Lee, C.~H.\ 2019, \apj, 875, 121
\bibitem[Lee(2020)]{Lee2020}Lee, C.~H., 2020, Astronomische Nachrichten, 341, 651
\bibitem[Leloudas et al(2015)]{Leloudas2015}Leloudas, G., et al.\ 2015, \apj, 815, L10
\bibitem[Leloudas et al(2017)]{Leloudas2017}Leloudas, G., et al.\ 2017, \apj, 837, L14
\bibitem[Lipunov et al.(1997)]{lipunov97} Lipunov, V.~M., Postnov, K.~A., \& Prokhorov, M.~E.\ 1997, \mnras, 288, 245
\bibitem[Lucy(1999)]{Lucy99}Lucy, L. B.\ 1999, \aap, 345, 211
\bibitem[Matzner \& McKee(1999)]{matzner99} Matzner, C.~D. \& McKee, C.~F.\ 1999, \apj, 510, 379
\bibitem[Maundet et al(2019)]{Maund2019}Maund, J. R., Steele, I., \& Jermak H.\ 2019, \mnras,
482, 4057
\bibitem[Maund et al(2020)]{Maund2020}Maund, J. R., Leloudas G., \& Malesani D. B.\ 2020, \mnras, 498, 3730
\bibitem[Maund et al(2021)]{Maund2021}Maund, J. R., et al.\ 2021, \mnras, 503, 312




\bibitem[Mazzali \& Lucy(1993)]{Mazzali93}Mazzali, P. A., \& Lucy, L. B. 1993, \aap, 279, 447
\bibitem[Page \& Thorne(1974)]{page74} Page, D.~N. \& Thorne, K.~S.\ 1974, \apj, 191, 499
\bibitem[Portegies Zwart \& McMillan(2000)]{portegies00} Portegies Zwart, S.~F., \& McMillan, S.~L.~W.\ 2000, \apjl, 528, L17
\bibitem[Rodriguez et al.(2015)]{rodriguez15} Rodriguez, C.~L., Morscher, M., Pattabiraman, B., et al.\ 2015, \prl, 115, 051101

\bibitem[Saito et al(2020)]{Saito2020}Saito S., et al.\ 2020, \apj, 894, 154
\bibitem[Shakura \& Sunyaev(1973)]{shakura73} Shakura, N.~I., \& Sunyaev, R.~A.\ 1973, \aap, 500, 33
\bibitem[Shapiro \& Sutherland(1982)]{Shapiro82} Shapiro, P. R., \& Sutherland, P. G.\ 1982, \apj, 263, 902
\bibitem[Sigurdsson \& Hernquist(1993)]{sigurdsson93} Sigurdsson, S., \& Hernquist, L.\ 1993, \nat, 364, 423
\bibitem[Soker(2022)]{Soker22}Soker, N.\ 2022, \araa, 22,122003

\bibitem[Strubbe \& Quataert(2009)]{strubbe09} Strubbe, L.~E. \& Quataert, E.\ 2009, \mnras, 400, 2070

\bibitem[Tanaka et al.(2017)]{Tanaka2017}Tanaka, M., Maeda, K., Mazzali, P. A., Kawabata, K. S., \& Nomoto, K. 2017, ApJ, 837, 105
\bibitem[Tutukov \& Yungelson(1973)]{tutukov73} Tutukov, A., \& Yungelson, L.\ 1973, Nauchnye Informatsii, 27, 70

\bibitem[Wang \& Wheeler(2008)]{Wang08} Wang, L., \& Wheeler, J.~C.\ 2008, \araa, 46, 433
\bibitem[Whitney(2011)]{Whitney2011} Whitney B. A.\ 2011, Bulletin of the Astronomical Society of India, 39, 101
\bibitem[Whitney(1992)]{Whitney92}Whitney, B. A., \& Hartmann, L.\ 1992, \apj, 395, 529



\end{thebibliography}

\appendix

\section{Three-dimensional Monte Carlo polarization simulation code}

We have developed a 3D Monte Carlo polarization code to calculate the continuum polarization spectrum and the line polarization spectrum for arbitrary 3D supernova envelope.
The Monte Carlo method used in this code is a common method for calculating polarization through the scattering process \citep{Daniel1980,Whitney92,Hillier91,Code95}, which is widely applied to the studies on SNe \citep{Hoflich91,Hoflich95,Mazzali93,Kasen03,Bulla15,Tanaka2017}.

\subsection{Setting grid}
Firstly, before the simulation, we need to discretize the computational domain into a grid. We set up a three-dimensional Cartesian space grid with a $100 \times 100 \times 100$ grid, and the parameters (e.g. density and electron number density) in the ejecta are constant in each grid. We assume a density profile of SN ejecta following a broken power law as described in Section 2.1  \citep{matzner99}.
Considering that the SN ejecta are homologous expanding ($r \propto v$), the spatial coordinates are determined by the radius of each ejecta layer. The outer boundary coordinates $R_{\rm max}$ of the ejecta correspond to the ejeta's maximum velocity $v_{\rm max}$. When the code calculates an arbitrary line with a rest wavelength of $\lambda_0$, we consider the range of wavelengths between
$\lambda_0 (1-v_{\rm max}/c)$ and $\lambda_0 (1+v_{\rm max}/c)$.
The energy spectrum is assumed to be constant in this wavelength range.

\subsection{Energy packet initialization}
Next, we need to initialize the photon package parameters. The N photon packets are emitted on a defined boundary (in this paper photon packets are emitted from the photosphere) and solve radiation transfer by tracking the photon packets propagating in the expanding ejecta. Each photon packet has a certain energy, wavelength, and Stokes parameters. In particular, each photon packet in the simulation has a constant energy, independent of the wavelength of the photon packet.
This treatment is consistent with \cite{Mazzali93,Lucy99,Kasen03,Bulla15}. In the case of this paper, the photons emitted from the two regions divided with $\theta_{\rm BP}$ on the photosphere carry different levels of energy, and the energy carried depends mainly on the luminosity of the emitting region. The luminosity of the emitting region ($L_{ \rm region \uppercase\expandafter{\romannumeral1}},~L_{\rm region \uppercase\expandafter{\romannumeral2}}$) and the energy ($\epsilon_{\rm\uppercase\expandafter{\romannumeral1}},~\epsilon_{\rm\uppercase\expandafter{\romannumeral2}}$) carried by the photon packet can be given by

\begin{equation}
L_{\rm region \uppercase\expandafter{\romannumeral1}} = \frac{N_{\rm \uppercase\expandafter{\romannumeral1}} \epsilon_{\rm \uppercase\expandafter{\romannumeral1}}}{\Delta t}, ~~
L_{\rm region \uppercase\expandafter{\romannumeral2}} = \frac{N_{\rm \uppercase\expandafter{\romannumeral2}} \epsilon_{\rm \uppercase\expandafter{\romannumeral1}}}{\Delta t},
\end{equation}
where $N_{\rm \uppercase\expandafter{\romannumeral1}}$ and $N_{\rm  \uppercase\expandafter{\romannumeral2}}$ are the number of photon packets emitted from region \uppercase\expandafter{\romannumeral1} and region \uppercase\expandafter{\romannumeral2} respectively and satisfy $N_{\rm \uppercase\expandafter{\romannumeral1}}+N_{\rm \uppercase\expandafter{\romannumeral2}}=N$. We consider a simple way to determine $N_{\rm \uppercase\expandafter{\romannumeral1}}$ and $N_{\rm \uppercase\expandafter{\romannumeral2}}$, for a photon packet take a random number $\theta$ satisfying $\theta\in[0, 180]$, if $\theta<\theta_{\rm BP}$ the photon packet is added to $N_{\rm \uppercase\expandafter{\romannumeral1}}$, otherwise added to $N_{\rm \uppercase\expandafter{\romannumeral2}}$. The wavelength information of each photon packet is sampled by assuming the photosphere surface to be a black body, where the photosphere surface temperature is determined by the ratio of the luminosity and area of the emitted region, as shown in Equation \ref{eq:Temp}.

The boundary position of the emitted photon packages is determined such that the electron scattering optical depth from the boundary to infinity is $\tau_{\rm in}$.
In the simulation process, it is generally used $\tau_{\rm in} = 3$ as adopted in \citet{Kasen03} and \citet{Hole10}. In this paper we consider $\tau_{\rm in} = \tau_{\rm ph} = 2/3$, which is the photon packet emitted from the photosphere as treated in \citet{Inserra2016}.
We assume that the photon packet emitted from the photosphere is unpolarized and the Stokes vector can be expressed as
\begin{equation}
I = 
\left( \begin{array}{c}
I\\ Q\\ U
\end{array} \right)
= 
\left( \begin{array}{c}
1\\ 0\\ 0
\end{array} \right).
\end{equation}
The initial emission direction of the photon is determined by 
$\mu = \sqrt{z}$ \citep{Mazzali93}
($z$ is a random number, $0 < z \leq 1$), 
where $\mu$ is cosine of the angle 
between the radial and photon direction.
The azimuthal angle around the radial direction $\psi$
is uniformly distributed, $\psi = 2 \pi z$.

\subsection{Simulation }
The emitted photon packets may experience three possible events during their propagation through the ejecta:
(1) escaping from the grid, 
(2) the electron scattering,
and (3) the line scattering \citep{Mazzali93}.
The actual occurs event is determined by calculating the length to 3 events. The photon packet is assigned a random optical depth $\tau = -\ln(z)$ ($0 < z \leq 1$), and an electron scattering event occurs when it reaches this optical depth during propagation. The distance to the electron scattering $l_{\rm elec}$ can be computed by $\tau = n_{e}(r) \sigma l_{\rm elec}$. With a certain position and direction vector of the photon packet the length $l_{\rm grid}$ to the next grid can be calculated. When $l_{\rm elec}$ is shorter than $l_{\rm grid}$ without considering line scattering, electron scattering occurs otherwise the photon package escapes the outer boundary of the ejecta.

The line scattering causes the photon packet to depolarize and re-emit along a new propagation direction with the same co-moving frame frequency. In the calculation of line interactions, we use the Sobolev approximation, which is a valid approximation in supernova ejecta with large velocity gradients \citep{castor70}. The distance to the line scattering event (Sobolev point) is  $l_{\rm line} = c (\lambda_0 - \lambda')/\lambda_0$,
where $\lambda'$ is the comoving wavelength of the photon packet.
If $l_{\rm line}$ is the shortest compared to $l_{\rm elec}$ and $l_{\rm grid}$,  the photon packet occurs line scattering interaction is considered.
The line scattering event actually occurs when 
the sum of the line scattering optical depth ($\tau_{\rm line}$) 
and the electron scattering optical depth $\tau_{\rm elec'}$ during the length $l_{\rm line}$
($\tau_{\rm elec'} = n_{e}(r) \sigma l_{\rm line}$) exceeds $\tau$.
If this sum does not reach $\tau$, then the opacity of the electron scattering is evaluated again and the photon packet eventually experiences electron scattering or escapes the grid.
For more details of this process refer to \cite{Mazzali93}.

Electron scattering events change the polarization properties of 
the photon packets. The phase matrix in the scattering frame can be written as follows \citep{chandrasekhar1960};
\begin{equation}
{R}(\Theta) = \frac{3}{4}
\left( \begin{array}{ccc}
\cos^2 \Theta + 1 & \cos^2 \Theta - 1 &  0 \\
\cos^2 \Theta - 1 & \cos^2 \Theta + 1 &  0 \\
0                 &        0          &  2 \cos \Theta
\end{array} \right),
\end{equation}
where $\Theta$ is the scattering angle on the scattering plane.
The rotation matrix for the Stokes parameters
is written as follows \citep{chandrasekhar1960};
\begin{equation}
{L}(\phi) =
\left( \begin{array}{ccc}
   1    &   0        &   0        \\
   0    & \cos 2\phi & \sin 2 \phi \\
   0    & -\sin 2\phi & \cos 2 \phi \\
\end{array} \right).
\end{equation}
Through the rotation matrix and the phase matrix, the Stokes parameters after electron scattering action is given by
\begin{equation}
{I}_{\rm out} = {L}(\pi - i_2) R(\Theta) {L}(-i_1) {I}_{\rm in}.
\label{eq:pol_escat}
\end{equation}
Where ${I}_{\rm in}$ and ${I}_{\rm out}$ is Stokes parameter
in the rest frame before and after the scattering, respectively.
The angles $i_1$ and $i_2$ are the angles on the 
spherical triangle defined as in \citep{chandrasekhar1960}.
Following \cite{Code95}, the scattering angle $\Theta$ of the electron scattering is chosen by sampling the probability distribution function ($f_{\rm pdf}$)
\begin{equation}
f_{\rm pdf} = \frac{1}{2} (\cos^2 \Theta + 1) + \frac{1}{2} (\cos^2 \Theta -1)
(\cos 2i_1 Q_{\rm in}/I_{\rm in} - \sin2i_1 U_{\rm in}/I_{\rm in}).
\end{equation}

The scattering event changes the energy and wavelength of the photon packet. The energy of the photon packet follows energy conservation in the rest frame,
\begin{equation}
\epsilon_{\rm out} = \epsilon_{\rm in} \frac{1 - \Vec{n}_{in} \cdot \Vec{v}/c}{1 -\Vec{n}_{out} \cdot \Vec{v}/c},
\end{equation}
where $\epsilon_{\rm in}$ and $\epsilon_{\rm out}$
are the rest-frame energy of the incoming and outgoing packets, respectively.
Similarly, the change in the wavelength is given by
\begin{equation}
\lambda_{\rm out} = \lambda_{\rm in} \frac{1 - \Vec{n}_{out} \cdot \Vec{v}/c}{1 - \Vec{n}_{in} \cdot \Vec{v}/c},
\end{equation} 
where $\lambda_{\rm in}$ and $\lambda_{\rm out}$ are the rest-frame 
wavelength of the incoming and outgoing packet, respectively.

\subsection{Test code}

\begin{figure}[tbph]
\begin{center}
\includegraphics[width=0.49\textwidth,angle=0]{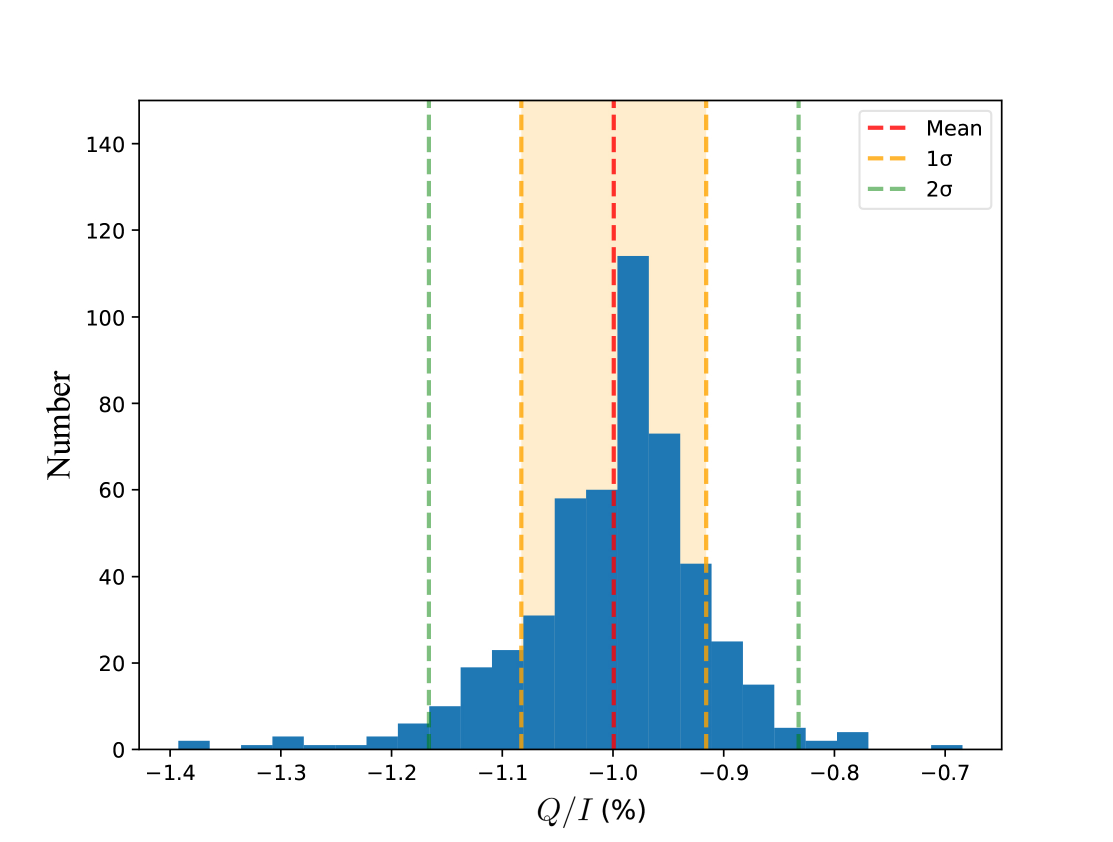}

\end{center}
\caption{The distribution of $P$ along the equatorial direction obtained from 500 simulations. The yellow and green dashed lines represent the $1\sigma$ and $2\sigma$ uncertainty ranges, respectively.}
\label{fig:A2}
\end{figure}

\begin{figure}[tbph]
\begin{center}
\includegraphics[width=0.49\textwidth,angle=0]{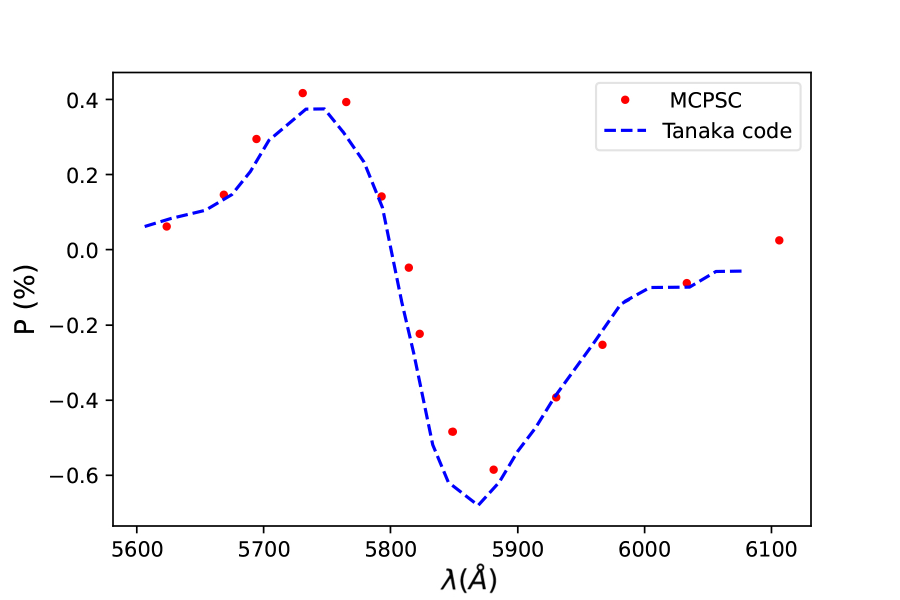}

\end{center}
\caption{The polarization spectrum with a torus clump distributed in the equatorial direction outside the photosphere. The line of sight is chosen to be 60 deg from the pole. The dashed blue line is the result predicted by \cite{Tanaka2017}, and the red dots are the results from our MCPSC.}
\label{fig:A1}
\end{figure}

Finally we need to verify the validity of the code. 
We did some simple tests and found our results for electron scattering are consistent with the results described in \cite{Code95} in the scenario of an optically thick blob. 
In order to assess the accuracy and uncertainty of the continuum polarization degree for a specific configuration, we conducted simulations of the continuum polarization signal generated by a prolate ellipsoidal envelope, using the parameter setup previously employed by  \citet{Inserra2016}. The simulation was run 500 times (N=500), and the distribution of the $P$ is presented in Figure \ref{fig:A2}. For each simulation, $10^6$ packets were generated under the assumption of a pure scattering atmosphere. The results are consistent with those reported by  \citet{Inserra2016}. and demonstrate a tight range of uncertainty.
For line scattering, our code can reproduce the H Lyman $\alpha$-line profile predicted by \cite{Jeffery1990}. To ensure the code is applicable to supernovae, we repeated the computation procedure in \cite{Tanaka2017} with the existence of ring-shaped clumps outside the SN photosphere. We only take a line with $\lambda=6000 \mathring{A}$. Figure \ref{fig:A1} shows our results (red dots) and the results in \cite{Tanaka2017} (blue line) under the same configuration. It can be seen that our results are consistent with theirs.

\end{document}